\documentclass[aps,prb,reprint,groupedaddress,superscriptaddress,floatfix]{revtex4-2}

\usepackage[T1]{fontenc}
\usepackage{newtxtext}
\usepackage{newtxmath}

\usepackage{amsmath,tikz,calc}
\usepackage{mathtools}
\usepackage{bm}
\usepackage{xfrac}
\usepackage{braket}

\usepackage{color}
\usepackage{xcolor}
\usepackage[colorlinks,
            citecolor=blue,
            linkcolor=blue,
            urlcolor=blue]{hyperref}

\usepackage{graphicx}
\usepackage[all]{hypcap} 
\usepackage{lipsum}

\usepackage{layout}
\usepackage{bm}   
\usepackage{bbold}
\usepackage{dcolumn}
\usepackage{braket}
\usepackage{float}

\begin{document}

\preprint{APS/123-QED}

\title{Multipartite entanglement in the 1-D spin-$\frac{1}{2}$ Heisenberg Antiferromagnet}

\author{Varun Menon}
\email{varunmenon@berkeley.edu}
\affiliation{
Department of Physics, University of California, Berkeley, Berkeley, CA 94720, USA
}%
\author{Nicholas E. Sherman}%
\affiliation{
Department of Physics, University of California, Berkeley, Berkeley, CA 94720, USA
}
\affiliation{Materials Sciences Division, Lawrence Berkeley National Laboratory, Berkeley, CA 94720, USA}
\author{Maxime Dupont}%
\affiliation{
Department of Physics, University of California, Berkeley, Berkeley, CA 94720, USA
}
\affiliation{Materials Sciences Division, Lawrence Berkeley National Laboratory, Berkeley, CA 94720, USA}


\author{Allen O. Scheie}
\affiliation{
MPA-Q, Los Alamos National Laboratory, Los Alamos, NM 87545, USA
}%

\author{D. Alan Tennant}
\affiliation{
Department of Physics and Astronomy, University of Tennessee, Knoxville, TN 37996, USA
}%
\affiliation{
Department of Materials Science and Engineering, University of Tennessee, Knoxville, TN 37996, USA
}%
\affiliation{
Shull Wollan Center, Oak Ridge National Laboratory, Oak Ridge, TN 37831, USA
}

\author{Joel E. Moore}%
\affiliation{
Department of Physics, University of California, Berkeley, Berkeley, CA 94720, USA
}
\affiliation{Materials Sciences Division, Lawrence Berkeley National Laboratory, Berkeley, CA 94720, USA}

\date{\today}

\begin{abstract}
 Multipartite entanglement refers to the simultaneous entanglement between multiple subsystems of a many-body quantum system. While multipartite entanglement can be difficult to quantify analytically, it is known that it can be witnessed through the Quantum Fisher information (QFI), a quantity that can also be related to dynamical Kubo response functions. In this work, we first show that the finite temperature QFI can generally be expressed in terms of a static structure factor of the system, plus a correction that vanishes as $T\rightarrow 0$. We argue that this implies that the static structure factor witnesses multipartite entanglement near quantum critical points at temperatures below a characteristic energy scale that is determined by universal properties, up to a non-universal amplitude. Therefore, in systems with a known static structure factor, we can deduce finite temperature scaling of multipartite entanglement and low temperature entanglement depth without knowledge of the full dynamical response function of the system. This is particularly useful to study 1D quantum critical systems in which sub-power-law divergences can dominate entanglement growth, where the conventional scaling theory of the QFI breaks down. The 1D spin-$\frac{1}{2}$ antiferromagnetic Heisenberg model is an important example of such a system, and we show that multipartite entanglement in the Heisenberg chain diverges non-trivially as $\sim \log(1/T)^{3/2}$. We verify these predictions with calculations of the QFI using conformal field theory and matrix product state simulations. Finally we discuss the implications of our results for experiments to probe entanglement in quantum materials, comparing to neutron scattering data in KCuF$_3$, a material well-described by the Heisenberg chain.
\end{abstract}

\maketitle


\section{\label{intro}Introduction}
Entanglement is one of the most celebrated and defining hallmarks of quantum theory and has become a central feature of modern physics in the age of quantum information, with a recent Nobel prize being awarded for the demonstration of quantum entanglement and violations of Bell inequalities \cite{nobel1, nobel2}. In the study of many-body quantum systems, the perspective of entanglement as an emergent property of interacting quantum degrees of freedom has become invaluable to our understanding of phase transitions, critical phenomena, and many-body dynamics. Universal scaling laws for entanglement growth at quantum critical points (QCPs) \cite{ee_scaling,calabrese_cardy,vidal_kitaev,Refael_Moore} have driven our understanding of quantum criticality and the identification of novel phenomena specific to many-body quantum systems, such as many-body localization \cite{mbl}, eigenstate thermalization \cite{eth1,eth2}, topological phases \cite{topo}, and universal quantum quench dynamics \cite{quench}, to name a few.\\

Entanglement is most commonly quantified through the entanglement entropy \cite{ee}, which measures quantum correlations between two subsystems in a bipartition of a larger system. Simultaneous entanglement between more than two subsystems, a phenomenon known as \textit{multipartite entanglement}, is more difficult to quantify. Multipartite entangled states can exhibit different degrees of separability, and can host stronger quantum correlations than the entanglement entropy captures \cite{multi_measures1}. While measures of multipartite entanglement with information-theoretic properties similar to the entanglement entropy have been proposed \cite{multi_measures1,multi_measures2,multi_measures3, EW}, such measures are highly non-local and intractable to compute for large systems in practice. Thus, understanding multipartite entanglement in many-body systems, particularly how it behaves in different phases and scales at quantum critical points, is of significant theoretical interest. On the other hand, the rapidly growing fields of quantum computing, communications, and sensing, bring with them a perspective of viewing entanglement as a computational resource. This perspective has driven efforts to develop quantum materials that host ground states with many-body entanglement that is robust to experimental conditions, for example, to make cat-state qubits \cite{cat} for quantum computers or probe states for quantum sensors \cite{sensing}. Probing entanglement measures such as the entanglement entropy requires knowledge of the spectrum of the density matrix of a system, which is computationally expensive for the complex ground states of quantum materials. There is therefore a strong desire to connect quantities measured in standard techniques for characterizing quantum materials, such as neutron scattering \cite{neutron_review} and NMR spectroscopy \cite{NMR_review}, with measures of multipartite entanglement. \\

A first step towards understanding multipartite entanglement with a local measure was provided independently by Tóth \cite{Toth} and Hyllus \textit{et al.} \cite{hyllus}, who showed that the Quantum Fisher Information (QFI), a measure originally from the field of quantum metrology that quantifies the sensitivity of a state to an unknown parameter \cite{toth2}, acts as a witness of multipartite entanglement -- an entanglement witness is a functional of a state which takes values that distinguish between sets of states with different degrees of entanglement \cite{EW}. Although the QFI is not an entanglement monotone \cite{Horodecki}, it has certain properties that make it an attractive measure in addition to its ability to distinguish multipartiteness of entanglement. The QFI witnesses entanglement in highly mixed or thermal states, unlike the entanglement entropy which degenerates to the classical Shannon entropy, dominated by classical correlations. As such, the QFI is a measure of pure quantum fluctuations, and is insensitive to fluctuations at thermal phase transitions \cite{hauke}, making it a particularly good probe for quantum criticality. The QFI has since been used to study multipartite entanglement in various many-body phenomena such as topological phase transitions \cite{qfi_topo}, many-body localization \cite{MBL_QFI}, and eigenstate thermalization \cite{eth_qfi}. Moreover, Hauke \textit{et al.} \cite{hauke} demonstrated a general relationship between the QFI and dynamical Kubo response functions of certain operators. Since dynamical response functions can be measured experimentally through neutron scattering experiments, their results provide a method for measuring multipartite entanglement in quantum materials, as demonstrated in recent experiments by Scheie \textit{et al.} and Laurell \textit{et al.}  \cite{scheie,scheie3}. \\

As a quantity sensitive to quantum fluctuations in thermal states, understanding how the QFI scales with finite temperature near a quantum critical point is important to both theoretical efforts to determine entanglement scaling laws, as well as to experiments that aim to detect entanglement in quantum materials. To this end, Hauke \textit{et al.} proposed a power-law for universal scaling of the QFI at quantum critical points \cite{hauke}. However, it is possible for the scaling exponent in this power-law to be zero in some critical systems. In this case, the scaling theory in Ref. \onlinecite{hauke} breaks down, and does not reveal possible logarithmic corrections that could dominate entanglement growth in the critical regime. An important example of such a system is the critical spin-$\frac{1}{2}$ antiferromagnetic Heisenberg model in one dimension. Aside from being a quintessential model of quantum criticality, the Heisenberg chain is a test-bed for quantum magnetism, describing many real antiferromagnetic materials. For example, it is well known that the quasi one-dimensional material KCuF$_3$ is well described by the spin-$\frac{1}{2}$ Heisenberg chain above $T \approx 40$ K 
\cite{kcuf3_og}. Recent experiments by Scheie \textit{et al.} have demonstrated up to 4-partite entanglement in KCuF$_3$ and the XXZ chain material $\text{Cs}_2\text{CoCl}_4$ through neutron scattering measurements of the QFI \cite{scheie_kcuf3, scheie3} \footnote{As of December 2022, references \cite{scheie_kcuf3,scheie3} state up to bipartite entanglement. Errata have been submitted by the original authors that clarify the correct form of the fluctuation-dissipation theorem to use in computing the QFI from neutron scattering data, which implies twice the entanglement depth than previously expected.}. However, a theoretical understanding of multipartite entanglement in the Heisenberg chain at finite temperature is still lacking. \\

One objective of this work is to study multipartite entanglement in the Heisenberg chain using analytical and numerical techniques, and to establish finite temperature scaling laws that agree with existing experimental data. To this effect, in section \ref{heisenberg_finite_T}, we show that the QFI density scales as
\begin{equation}\label{QFI_scaling_T}
    f_Q \sim \log\left(\frac{1}{T}\right)^{3/2}
\end{equation}
in the asymptotic limit $T\rightarrow 0$. To prove \eqref{QFI_scaling_T}, we also derive general results that relate multipartite entanglement to static structure factors of certain operators. In particular, in section \ref{analytical_argument} we show that the QFI can be expressed as a quantity proportional to the static structure factor of an operator, plus a temperature dependent correction term that vanishes as $T\rightarrow 0$. We argue that this implies that the static structure factor witnesses multipartite entanglement below a threshold temperature, indicating that quantum fluctuations dominate the structure factor below a characteristic energy scale. In section \ref{CFT}, we apply these results to analyze multipartite entanglement in the Heisenberg chain from low-energy conformal field-theoretic expressions of spectral functions \cite{starykh1, starykh2}. We further verify the predicted scaling in \eqref{QFI_scaling_T} with a matrix product state (MPS) approach \cite{schollwock_mps} in section \ref{mps}. Finally, in section \ref{experiment} we compare our predictions to neutron scattering data for KCuF$_3$, and discuss experimental implications of our results, suggesting candidate systems for future experiments to detect diverging multipartite entanglement at non-zero temperature.
\section{\label{technical_review}Technical Background}
We review the definition of multipartite entanglement in terms of a separability hierarchy, the Quantum Fisher Information and its expression in terms of dynamical response functions, as well as its real-space renormalization scaling theory.
\subsection{\label{multipartite_entanglement}Multipartite Entanglement}
We define multipartite entanglement as in \cite{hyllus, Toth, guhne_and_toth1,seevinck,chen,guhne}. An $N$-particle pure state $\rho$ is \textit{k-separable}, if it can be written as the tensor product of factor states of not more than $k$ particles each. That is, $\rho$ can be expressed as
\begin{equation*}
 \rho = \bigotimes_{i=1}^M \rho^{K_i}
\end{equation*}
where $\rho^{K_i}$ are states on disjoint subsets $K_i \subset \{1\cdots N\}$ of no more than $k$ particles each. \\
A mixed state $\rho$ is $k$-separable if it can be written as a mixture of $k$-separable pure states. That is, $\rho = \sum_i\lambda_i \rho_i$ where each $\rho_i$ is $k$-separable and $\lambda_i$ is the probability of the i\textsuperscript{th} state in the mixture. \\

In either case, $\rho$ is \textit{k-partite entangled} if and only if it is $k$-separable but not $k-1$-separable. This is a direct generalization of the definition of bipartite entanglement to higher orders of infactorability of a many body density matrix. In this work, we also use the term \textit{entanglement depth} of a quantum state, which is the largest $k$ for which the state is $k$-partite entangled.

\subsection{\label{QFI_review}Quantum Fisher Information}
Analogous to the classical Fisher information \cite{fisher}, the Quantum Fisher Information (QFI) was initially developed as a measure of the statistical sensitivity of a quantum state to a unitarily encoded parameter \cite{qfi_original_paper,toth2,liu}. The QFI is a functional of a quantum state $\rho$ and an operator $\mathcal{O}$. For a pure state, the QFI reduces to a quantity proportional to the variance of the operator $\mathcal{O}$ in the state $\rho$
\begin{equation}\label{ground_qfi}
    F_Q[\rho, \mathcal{O}] = 4\text{Var}(\mathcal{O}) = 4\left(\left<\mathcal{O}^2\right> - \left<\mathcal{O}\right>^2 \right)
\end{equation}

For a mixed state, the QFI is a generalization of the variance that captures the quantum, but not classical, fluctuations in the operator $\mathcal{O}$, defined for a mixed state $\rho = \displaystyle\sum_i \lambda_i \ket{\psi_i}\bra{\psi_i}$ as 
\begin{equation}\label{qfi_def}
F_Q[\rho, \mathcal{O}] = 2\sum_{i\neq j} \frac{(\lambda_i - \lambda_j)^2}{\lambda_i+\lambda_j}|\bra{\psi_i}\mathcal{O}\ket{\psi_j}|^2
\end{equation}
where $\ket{\psi_i}$ is the i\textsuperscript{th} eigenstate of the mixed state $\rho$ with eigenvalue $p_i$ \cite{qfi_original_paper}. 

\subsubsection{\label{QFI_entanglement}QFI and multipartite entanglement}
We consider observables that are total spin operators $\mathcal{O} = \displaystyle\sum\limits_i e^{i\phi_i} \vec{S}^z_i$, where $\vec{S}_i^z = \frac{1}{2} \vec{\sigma_i^z}$ is the z-component spin operator at site $i$, and $e^{i\phi_i}$ are phases. We use $S^z$ because the Heisenberg model is rotationally invariant \cite{faddeev}. For spin models without $SU(2)$ symmetry, the spin direction that maximizes the QFI is an appropriate choice \cite{QFI_XY}. Then for an $N$-particle state $\rho$, if 
\begin{equation}\label{qfi_bound}
    f_Q = \frac{F_Q}{N} > k
\end{equation}
for $k$ a divisor of $N$, then $\rho$ is at least $k+1$-partite entangled \cite{hyllus,Toth,hauke}\footnote{This bound may be refined to include $k$ not a divisor of $N$ as shown in Ref. \cite{hauke}.}. Note that \eqref{qfi_bound} is a one way implication, that is, $f_Q < k$ does not imply less than $k+1$ partite entanglement in $\rho$. \\
\subsubsection{\label{QFI_hauke_section} QFI of thermal states from dynamical response functions}
The QFI for a thermal mixed state at inverse temperature $\beta = \frac{1}{T}$ can be expressed in terms of dynamical Kubo response functions as \cite{hauke}
\begin{equation}\label{qfi_integral}
    f_Q[\rho, \mathcal{O}, \beta] = \frac{4}{\pi}\int_0^\infty \mathrm{d}\omega \tanh\left(\frac{\beta\omega}{2}\right)\chi^{\prime\prime}(\omega, \beta)
\end{equation}
where $\chi^{\prime\prime}(\omega, \beta)$ is the imaginary (dissipative) part of the dynamic response function with respect to $\mathcal{O}$ in the state $\rho$ defined by the Kubo formula \cite{Kubo}
\begin{equation}\label{kubo}
    \chi(\omega, \beta) = \frac{\displaystyle i}{\displaystyle N}\int_0^\infty \mathrm{d}t \ e^{i\omega t} \mathrm{tr}\left(\rho \left[\mathcal{O}(t), \mathcal{O}\right] \right)
\end{equation} 
Due to the suppression of low $\omega$ contributions by the $\tanh(\frac{\omega}{2T})$ term, equation \eqref{qfi_integral} shows that high frequency quantum fluctuations in $\chi^{\prime\prime}(\omega,T)$ are the primary contributions to multipartite entanglement in the system \cite{hauke}. As $T\rightarrow 0 \ $, $\tanh(\frac{\omega}{2T}) \rightarrow 1$ for all $\omega$, and quantum fluctuations at all energy scales contribute to multipartite entanglement. 

\subsubsection{\label{QFI_hauke_scaling_section} Power-law scaling theory of the QFI}
Using real-space renormalization arguments, Ref. \cite{hauke} shows that at a 1-D quantum critical point, the QFI density scales as 
\begin{equation}\label{qfi_scaling_hauke}
    f_Q = \lambda^{\Delta_Q}\phi(T\lambda^z, L^{-1}\lambda, h\lambda^{\frac{1}{\nu}}) + c
\end{equation}
where $\phi$ is a universal function of its dimensionless arguments, $c$ is a non-universal constant, and $\lambda$ is the correlation length cutoff-scale associated with perturbations from the critical point. In Eq. \eqref{qfi_scaling_hauke}, $\nu$ and $z$ are the correlation-length critical exponent and dynamical critical exponent of the critical point respectively \cite{cardy_book}, $L$ is the system size, and $h$ is the strength of a conjugate field that drives the system away from criticality. The scaling exponent of the QFI is $\Delta_Q = 1-2\Delta_\mathcal{O}$, where $\Delta_\mathcal{O}$ is the scaling dimension of the operator $\mathcal{O}$ under a renormalization transformation \cite{cardy_book}. Equation \eqref{qfi_scaling_hauke} implies that critical points with $\Delta_Q  > 0$ host diverging multipartite entanglement as the system is brought closer to criticality, while for $\Delta_Q < 0$ multipartite entanglement asymptotes to a maximum value. For QCPs with conformal symmetry, Ref. \onlinecite{QFI_CFT} further shows that critical points that have bigger central charges and operators with smaller scaling dimensions have greater multipartite entanglement content, consistent with \eqref{qfi_scaling_hauke}. \\

Eq. \eqref{qfi_scaling_hauke} encodes how multipartite entanglement is expected to scale with finite temperature at a 1-D quantum critical point. At sufficiently low temperature $T$ such that the system is in the critical regime and non-zero temperature is the most relevant perturbation away from criticality, the correlation length of the system is cut off at a length scale $\lambda \sim T^{-\frac{1}{z}}$. Then, Eq. \eqref{qfi_scaling_hauke} implies that 
\begin{equation}\label{QFI_T_scaling_crude}
    f_Q(T)\sim T^{-\frac{\Delta_Q}{z}} 
\end{equation}
However, when $\Delta_Q = 0$, \eqref{qfi_scaling_hauke} and \eqref{QFI_T_scaling_crude} do not reveal possible sub-power-law terms that may dominate critical scaling.

\section{\label{analytical_argument}Finite temperature scaling of multipartite entanglement from spectral functions}
We consider operators that are sums of local operators at wavenumber $q$, \begin{equation}\label{total_op}
    \mathcal{O} = \sum\limits_{x=0}^{N} e^{iqx}O(x)
\end{equation} where $O(x)$ is a a local operator acting at site $x$. We also assume that $\left<O(x)\right> = 0$. These operators are typical order parameters of 1-D lattice models, sensitive to critical fluctuations in these systems, and are thus appropriate to compute the QFI with respect to as an entanglement witness \cite{hauke}.  

\subsection{\label{qfi_static_section}QFI and the Static Structure Factor}
We first show that the QFI of a thermal state can be expressed as 
\begin{align}\label{QFI_correction}
f_Q[\rho, \mathcal{O}, \beta] &= 4S(q, \beta) - 16\int_0^\infty d\omega \ \frac{S(q,\omega,\beta)}{1+e^{\beta\omega}} \nonumber\\
&= 4S(q, \beta)-\epsilon(\beta)
\end{align}
where $S(q)$ is the static structure factor with respect to $\vec{S}_{\text{total}}(q)$
\begin{equation}
    S(q) = \frac{\displaystyle 1}{\displaystyle N}\sum\limits_{x,y}e^{-iq(x-y)}\left<O(x)O(y)\right> 
\end{equation}
and $S(q,\omega)$ is the associated dynamic structure factor (DSF)
\begin{equation}\label{dsf}
    S(q,\omega) = \frac{1}{2\pi N}\int_{-\infty}^\infty \mathrm{d}t \sum\limits_{x,y}e^{-i(q(x-y)-\omega t)}\left<O(x)O(y)\right> 
\end{equation}
$S(q,\beta)$ is equivalent to the variance of $\mathcal{O}$ in a thermal state at inverse temperature $\beta$. Thus, the first term on the right hand side of \eqref{QFI_correction} is the finite temperature extrapolation of the expression for the ground state QFI, while $\epsilon(\beta)$ produces a temperature dependent correction that we show vanishes as $T\rightarrow 0$, reproducing \eqref{ground_qfi}. \\

To show \eqref{QFI_correction}, we first express the integral formulation of the QFI \eqref{qfi_integral} in terms of structure factors through the fluctuation dissipation theorem (see footnote \footnote{There are various versions of the factor relating $S(\omega)$ to $\chi^{\prime\prime}(\omega)$ depending on how each is defined. This is the correct form of the fluctuation-dissipation theorem to use when $\chi(\omega)$ is defined as an integral from $0$ to $\infty$ as in \eqref{kubo}, and $S(\omega)$ is not symmetrized in negative frequencies, so that the detailed balance condition $S(-\omega) = e^{-\beta\omega}S(\omega)$ holds, and $S(\omega)$ is in absolute units. See Ref. \cite{Lovesey1984}.}) \cite{recursion,Lovesey1984}
\begin{equation}\label{fluctuation_dissipation}
    \chi^{\prime\prime}(q,\omega) = \pi\left(1-e^{-\beta\omega} \right)S(q,\omega)
\end{equation}
so that 
\begin{eqnarray}\label{qfi_dsf}
    f_Q &= 4\displaystyle\int_0^\infty \mathrm{d}\omega \ \tanh\left(\frac{\beta\omega}{2}\right)(1-e^{-\beta\omega})S(\omega) \nonumber \\
        &= 8 \displaystyle\int_0^\infty \mathrm{d}\omega \ \frac{\displaystyle\cosh(\beta\omega) - 1}{\displaystyle e^{\beta\omega}+1} \cdot S(\omega)
\end{eqnarray}
where we drop the $q$ dependence when unambiguous. Next, we use the sum rule relating the static structure factor to the dynamic structure factor \cite{sum_rule}, 
\begin{eqnarray}\label{dsf_sum_rule}
    &S(q) = \displaystyle\int_{-\infty}^\infty \mathrm{d}\omega \ S(q,\omega) 
\end{eqnarray}
and the detailed-balance condition $S(-\omega) = e^{-\beta\omega}S(w)$ \cite{recursion}, so that
\begin{align}\label{correction}
    &f_Q - 4S(q) \nonumber \\
    &=8 \displaystyle\int_0^\infty \mathrm{d}\omega \ \left[ \frac{\displaystyle\cosh(\beta\omega) - 1}{\displaystyle e^{\beta\omega}+1}  -  \ \frac{\displaystyle(1+e^{-\beta\omega})}{\displaystyle 2}\right] S(\omega) \nonumber \\
    &=8 \displaystyle\int_0^\infty \mathrm{d}\omega \frac{(\cosh(\beta\omega)-1) - (\cosh(\beta\omega) +1)}{\displaystyle 1+e^{\beta\omega}} S(\omega) \nonumber \\
    &= -16 \displaystyle\int_0^\infty \mathrm{d}\omega \ \frac{S(\omega)}{\displaystyle 1+e^{\beta\omega}}
\end{align}
which proves \eqref{QFI_correction}. The coefficient of $S(\omega)$ in \eqref{correction} is a Fermi-Dirac distribution and is step-function-like as $\approx \Theta(T-\omega)$. For non-critical many-body systems, the dynamic structure factor $S(\omega)$ is finite and decays to $0$ rapidly for $\omega \gtrapprox \Delta$, where $\Delta$ is the spectral bandwidth of excitations created by the operator $\mathcal{O}$ \footnote{An operator that is a sum of local operators cannot create excitations of arbitrarily large energy. Thus, $\Delta$ is finite for such operators.}. As $T\rightarrow 0$ the Fermi-Dirac factor in the integral in \eqref{correction} reduces the effective domain of integration of $S(\omega)$ to $0$, which causes the term on the right side of \eqref{correction} to vanish, implying that $f_Q = 4S(q,T=0) = 4 \text{Var}(\vec{S}_{\text{total}}(q))$ as expected. At a quantum critical point, it is possible for $S(\omega)$ to be dominated by gapless excitations, so that the dynamic structure factor diverges at most algebraically at $T=0$  \cite{sachdev_senthil,chubukov, Muller}
\begin{equation*}
   S(\omega) \sim \omega^{-\alpha}
\end{equation*}

However, as $T\rightarrow 0$, the width of the Fermi-Dirac factor in \eqref{correction} vanishes exponentially fast, removing contributions from the divergence of $S(\omega)$ at $\omega=0$, implying once more that $f_Q = 4 \text{Var}(\vec{S}_{\text{total}}(q))$. Equation \eqref{correction} can be given the following physical interpretation: at $T=0$, the QFI is identically $4S(q)$, and quantum fluctuations at all energy scales contribute to the QFI through the sum rule \eqref{dsf_sum_rule}. As $T$ increases, thermal fluctuations introduce excitations into the static structure factor at higher energies. As the QFI is sensitive to purely quantum fluctuations that contribute to multipartite entanglement \cite{hauke}, these higher energy thermal excitations are removed by the Fermi-Dirac weight in the integral in \eqref{correction} to isolate quantum fluctuations in the spectral bandwidth of $S(\omega)$. This can also be interpreted as a decomposition of the variance of a thermal state into quantum and classical contributions in terms of spectral functions. The integral correction term in \eqref{correction} is then the variance of the operator that comes from thermal occupation probabilities.   \\

Equation \eqref{correction} also implies that all the information about the low temperature scaling of multipartite entanglement is contained in the static structure factor. In particular, since the right side of \eqref{correction} goes to $0$ as $T\rightarrow 0$, there exists some non-zero temperature $T_Q$ such that 
\begin{equation}\label{qfi_T_Q_bound}
   \epsilon(T_Q) = 4S(q,T_Q) - f_Q(T_Q) = 16\int_0^\infty \mathrm{d}\omega \ \frac{S(T_Q, \omega)}{\displaystyle 1+e^{\frac{\omega}{T_Q}}} < 1 
\end{equation}

Then, the relation \eqref{qfi_bound} between the QFI and multipartite entanglement implies that the static structure becomes an entanglement witness below $T < T_Q$, as 
\begin{equation}\label{qfi_static}
    4S(\rho, q, T) > k \Rightarrow \ \rho \textit{ at least k-partite entangled}
\end{equation}
for $k$ a divisor of N. Note that this differs by the bound in Eq. \eqref{qfi_bound} by a difference of 1. Therefore, calculating the static structure factor at sufficiently low temperature provides a path to determining multipartite entanglement content in the system. This approach has significant merits in practice, as it is generally more tractable to calculate static properties at finite temperature than the full spectrum of dynamic interactions or determining exact scaling dimensions of critical operators as required by \eqref{qfi_integral} or \eqref{qfi_scaling_hauke}. Furthermore, the static structure factor of operators of the form \eqref{total_op} can be measured in various quantum materials in neutron scattering \cite{static_scaling_exp1, static_scaling_exp2, static_scaling_exp3} and NMR \cite{static_scaling_nmr1,static_scaling_nmr2,static_scaling_nmr3} experiments with relative ease. \\

\subsection{\label{universality}\texorpdfstring{$T_Q$ as an energy scale}{}}

$T_Q$ can be interpreted as an energy scale below which $S(q)$ is dominated by quantum fluctuations and becomes an effective entanglement witness. We discuss the dependence of $T_Q$ on characteristic energy scales of the system.\\

In gapped systems with energy gap $\delta$, $S(q, \omega)$ has no spectral weight at $\omega \lessapprox \delta$. The Fermi-Dirac factor in \eqref{QFI_correction} weights frequencies below $T$, which implies that $4S(q) - f_Q \approx 0$ for $T\lessapprox \delta$, and becomes non-zero at higher temperatures. Thus, 
\begin{equation}
    T_Q \sim \delta
\end{equation}
in gapped systems, and $\delta$ is the only energy scale that determines $T_Q$. \\

The situation is different at 1-D quantum critical points where a gap closes. We assume the system possesses an overall energy scale $J$. For example, $J$ could be the coupling energy in a spin-chain. We assume $J=1$ for this discussion. Focusing on Lorentz invariant QCPs with dynamical critical exponent $z=1$ \cite{cardy_book, sachdev_book}, scale invariance implies a low temperature scaling form of the dynamic structure factor

\begin{equation}\label{S_omega_critical}
S(\omega) = \frac{A}{T^{(1-\eta)\nu}}\frac{\phi(\frac{\omega}{T})}{(1-e^{-\frac{\omega}{T}})}
\end{equation}

at fixed wavenumber $q$, where $\phi$ is a universal real function of $\frac{\omega}{T}$, $A$ is a non-universal dimensionless amplitude, $\nu$ is the correlation length critical exponent, and $\eta$ is the critical exponent associated with anomalous dimension \cite{sachdev_book,sachdev_senthil,chubukov}. This holds for $T\ll J$ and $\eta \neq 1$. At $T\sim \Delta$, where $\Delta$ is the spectral bandwidth of $S(\omega)$, the coefficient of $S(\omega)$ in the integral \eqref{qfi_dsf} is smaller than $1$ for $\omega < T \sim \Delta$. Comparing \eqref{qfi_dsf} and \eqref{dsf_sum_rule}, we then expect $T_Q \ll \Delta \sim J$, where the scaling in Eq. \eqref{S_omega_critical} holds. Inserting equation \eqref{S_omega_critical} into \eqref{correction}, at low temperatures we find
\begin{align}\label{universal_TQ}
    \epsilon(T) &= \frac{A}{T^{(1-\eta)\nu}}\int_0^\infty \mathrm{d}\omega \  \frac{\phi(\frac{\omega}{T})}{\sinh(\frac{\omega}{T})} 
    &= \frac{AD}{ T^{(1-\eta)\nu-1}}
\end{align}
where  
\begin{equation*}
    D = \displaystyle \int_0^\infty \mathrm{d}u \  \frac{\phi(u)}{\sinh(u)}
\end{equation*}
is a universal constant \footnote{D is a well defined number because $\frac{1}{\sinh(\frac{\omega}{T})}$ decays exponentially for $\omega/T > 1$. Thus, the integral in \eqref{universal_TQ} can be shown to converge by expanding $\phi$ as a power series in small $\omega/T$. }. From \eqref{qfi_T_Q_bound}, we deduce that
\begin{equation}\label{S_omega_critical3}
    T_Q \sim \displaystyle\left(AD\right)^{\frac{1}{(1-\eta)\nu-1}}
\end{equation}

for QCPs with $\eta \neq 1$. The theoretical implication of \eqref{S_omega_critical3} is that quantum fluctuations that contribute to multipartite entanglement become the dominant contribution to the static structure factor at a scale $T_Q$ determined by universal properties of the critical point, a non-universal amplitude, and the overall energy scale $J$. In particular, for a given model, the only energy scale that $T_Q$ depends on is $J$. However, for critical points with $\eta=1$, \eqref{S_omega_critical} breaks down and sub leading corrections to $S(\omega)$ that violate scale-invariance and introduce other energy scales could determine $T_Q$. The Heisenberg model is one such example, which we discuss in section \ref{CFT}.\\

In practice, $T_Q$ can be estimated from Eq. \eqref{correction} given a low energy approximation of the finite temperature DSF. Because of the Fermi-Dirac factor in \eqref{correction}, it suffices to use an approximation of $S(\omega, T)$ that is accurate at only low energies. Such estimates can often be made for spin chains using Bosonization and effective field theory methods \cite{calabrese2,bougorzi,starykh1, Tennant}.

\section{\label{heisenberg_finite_T} Finite temperature multipartite entanglement in the Heisenberg Chain}
We apply the ideas of section \ref{analytical_argument} to the spin-$\frac{1}{2}$ antiferromagnetic Heisenberg chain in one dimension, defined by the Hamiltonian
\begin{equation}\label{Heisenberg}
       H = J\sum_i S^x_iS^x_{i+1} + S^y_i S^y_{i+1} + S^z_iS^z_{i+1}
\end{equation}
where $S^\alpha_i = \frac{1}{2}\sigma^\alpha_i$ is the $\alpha$ component of the total spin operator at site $i$, $\alpha \in \{x,y,z\}$. We set $J=1$ for the rest of this work. The $T=0$ Heisenberg chain is the $SU(2)$ symmetric critical point that separates gapped and gapless phases of the XXZ model \cite{XXZ, faddeev}. 

\subsection{\label{heisenberg_scaling}Critical scaling of multipartite entanglement}
The power-law scaling theory of the QFI utilizes the scaling dimension $\Delta_{\mathcal{O}}$ of relevant operators. For the antiferromagnetic spin-$\frac{1}{2}$ Heisenberg chain, it is known that the most relevant operator is the staggered total spin operator $S^z_{\text{total}}(q=\pi) = \sum_i (-1)^iS_i^z$ \footnote{We consider $\alpha = z$ without loss of generality since the Heisenberg model is $SU(2)$ symmetric.} and has scaling dimension $\Delta = \frac{1}{2}$ \cite{scalingdimension}. Thus, the scaling exponent of the QFI in \eqref{QFI_T_scaling_crude} is $\Delta_Q = 0$, which does not reveal sub-power-law terms that contribute to multipartite entanglement. Our discussion in section \ref{analytical_argument} implies that at sufficiently low temperature, the static structure factor with respect to $S^z_{\text{total}}(q=\pi)$ witnesses multipartite entanglement through equations \eqref{correction} and \eqref{qfi_static}. The finite temperature staggered spin structure factor of the Heisenberg chain has been well studied analytically, numerically, and experimentally  \cite{starykh1,sachdev,static_scaling_exp1,static_scaling_exp2} and is known to display a power-logarithmic divergence, from general scaling arguments, as $T\rightarrow 0$ 
\begin{equation}\label{static_heisenberg_scaling}
    S(\pi, T) = D\left(\log\left(\frac{\displaystyle T_0}{\displaystyle T}\right)\right)^{3/2}
\end{equation}
for constants $D$ and $T_0$ \cite{starykh2}. Therefore, we expect multipartite entanglement in the Heisenberg chain to diverge at low temperatures, with critical scaling as in \eqref{static_heisenberg_scaling}, which reduces to \eqref{QFI_scaling_T} as $T\rightarrow 0$.

\subsection{\label{CFT} QFI from Conformal Field Theory}
We first estimate the temperature scale $T_Q$ below which the static structure factor witnesses multipartite entanglement through the bound \eqref{qfi_static}. At low energies, the Hamiltonian \eqref{Heisenberg} can be expressed in terms of an effective Tomonaga-Luttinger-Liquid (TLL) conformally invariant quantum field theory \cite{Haldane,Haldane2}. From the TLL model, Starykh, Singh, and Sandvik \cite{starykh1} proposed an analytical expression for the dynamical susceptibility $\chi(\omega, q)$ of the staggered spin operator in the Heisenberg chain, which extends the free-boson approach developed by Schulz \cite{Schulz} with multiplicative logarithmic corrections that violate universal $\frac{\omega}{T}$ scaling of $\chi(q,\omega)$. By adopting known finite-size scaling relations \cite{koma} to finite temperatures with conformal mappings of space-time correlation functions, their results imply the following low-energy expression: 
\begin{align}\label{starykh}
   &\chi^{\prime\prime}(\omega, q=\pi) = \frac{2^{2\Delta - \frac{3}{2}}D}{\pi T}\sin(2\pi\Delta) \left(\log\left(\frac{T}{T_0}\right) \right)^\frac{1}{2}  \nonumber\\
   &\displaystyle\times \text{Im}\left[\Gamma^2(1-2\Delta) \cdot \left(\frac{\Gamma\left(\Delta - i\cdot \frac{\omega}{4\pi T}\right)}{\Gamma\left(1 - \Delta - i\cdot \frac{\omega}{4\pi T}\right)}\right)^2\right]
\end{align}
where 
\begin{equation}
    \Delta = \frac{1}{4}\left(1-\frac{1}{2 \log\left(\frac{T_0}{T} \right)}\right)
\end{equation}
\begin{figure}
\includegraphics[width=8.6cm]{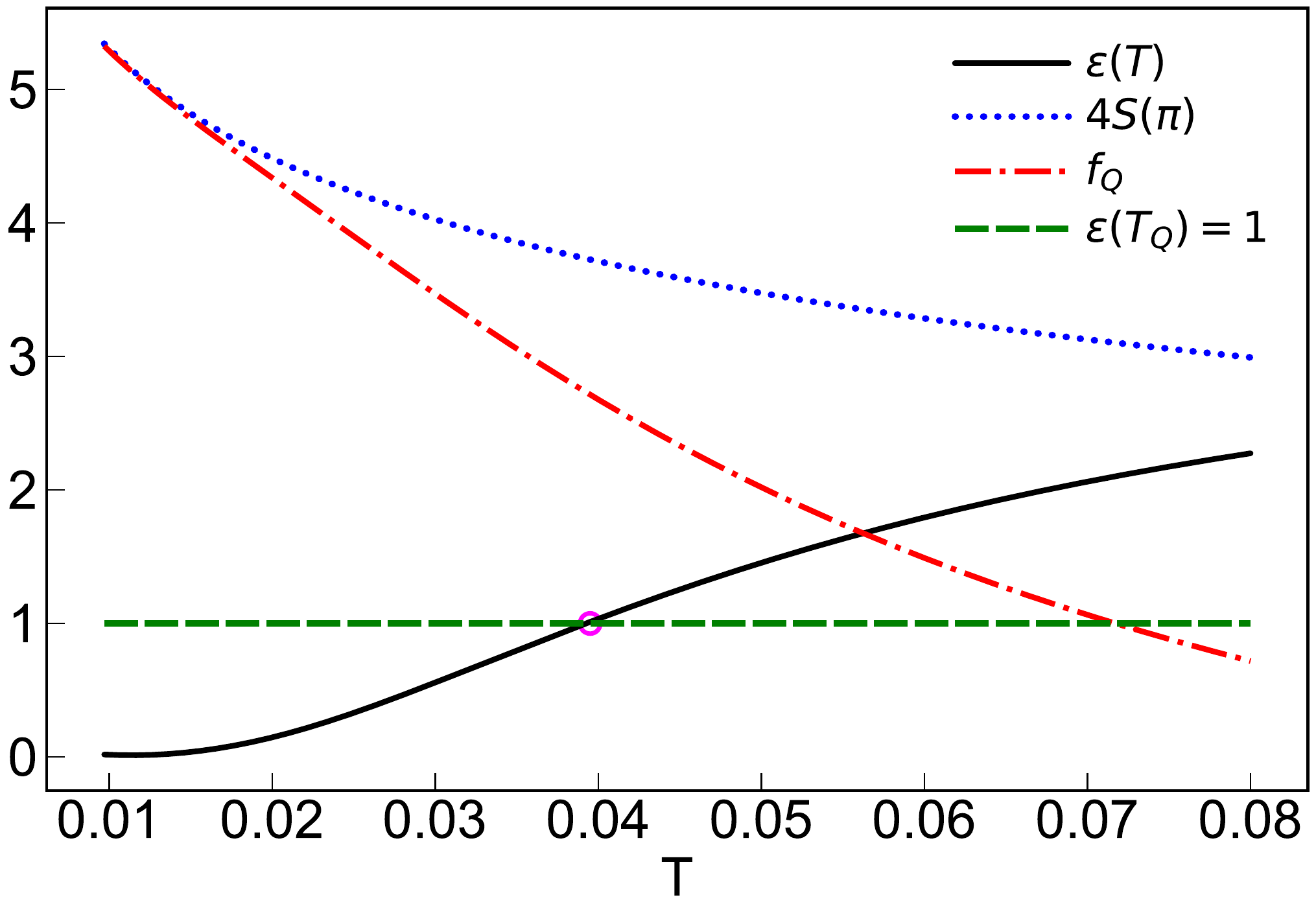}
\centering
\caption{\label{starykh_plot_T} $\epsilon = 4S(\pi) - f_Q = 16\int_0^\infty \mathrm{d}\omega \frac{S(\omega)}{1+e^{\omega/T}}$ as a function of temperature using Eq. \eqref{starykh}. The circle indicates where $4S(\pi) - f_Q$ becomes negligible at $T_Q \approx 0.04 J$, below which the static structure factor witnesses entanglement through Eq. \eqref{qfi_static}. $4S(\pi)$ approximated from the TLL model is also shown, along with $f_Q = 4S(\pi) - \epsilon$. At $T = 0.01 J$, at least $6-$partite entanglement is witnessed. }
\end{figure}
and $D$ and $T_0$ are effective constants in the low-temperature regime. Quantum Monte Carlo simulations in \cite{starykh1} provide the estimates $D=0.075$ and $T_0 = 4.5$, which have been demonstrated to agree well with experimental data \cite{takigawa}. Because \eqref{starykh} is derived from a low-energy continuum field theory, it fails to capture the effect of finite spectral bandwidth  due to finite lattice spacing, and overestimates contributions at large energies $\omega \gtrapprox \frac{1}{a}$ for lattice spacing $a$. Eq. \eqref{starykh} also becomes less accurate at $T\sim J$ as thermal fluctuations break the assumed linear dispersion of the TLL theory. However, the Fermi-Dirac weight in Eq. \eqref{correction} implies that at low temperature, contributions from high frequency components of $S(\omega)$ decay to $0$ rapidly, making the correction term $\epsilon = 4S(\pi) - f_Q$ from the low energy field theory an accurate approximation. In Fig. \ref{starykh_plot_T} we plot $\epsilon(T)$ using Eq. \eqref{correction} and Eq. \eqref{starykh}, and find that the finite temperature correction from the static structure factor becomes negligible at $T_Q\approx 0.04 J$ \footnote{A lower cutoff in $\omega$ and $T$ is set when computing the integral in \eqref{correction} to avoid numerical instabilities from the divergence of $S(\omega)$ as $\omega \rightarrow 0$ and $T\rightarrow 0$. Furthermore, the effective constants $D$ and $T_0$ in \eqref{starykh} are QMC estimates, and potentially have a slow temperature dependence which we neglect. Together, these effects introduce small errors into the estimate for $T_Q$. Nevertheless our analysis implies that on the order of $T\lessapprox 0.04 J$, the static structure factor closely approximates the QFI.}. Since $\eta=1$ for the Heisenberg antiferromagnet \cite{sachdev_book}, the scaling analysis of section \ref{universality} does not hold, and the energy scale $T_0$ introduced by the logarithmic correction to $S(\pi,\omega)$ in \eqref{starykh} also affects $T_Q$.\\

We can further determine the low-temperature entanglement depth in the Heisenberg chain using results from Bosonization. Starykh \textit{et al.} \cite{starykh1} also proposed an expression for the static structure factor using conformal mappings of equal time correlation functions to include finite temperature effects, which for $q=\pi$ reduces to

\begin{align}\label{starykh_static_structure_factor}
    S(q=\pi, T) &= 2^{\Delta+\frac{1}{2}}D\log\left(\frac{T_0}{T} \right)^{\frac{1}{2}} \nonumber\\
    &\displaystyle\times \Gamma(1-4\Delta)\cdot \mathrm{Re}\left(\frac{\Gamma(1-2\Delta)}{\Gamma(2\Delta)} \right)
\end{align}
which reproduces the expected $\left(\log\left(\frac{1}{T}\right)\right)^{\frac{3}{2}}$ scaling of $S(q=\pi, \omega)$ at low $T$. For $T < 0.04 J$, we expect $|\epsilon(T)| < 1$, so that $S(\pi)$ witnesses multipartite entanglement according to the bound \eqref{qfi_static}. In Fig. \ref{starykh_plot_T}, we plot $4S(\pi)$ from \eqref{starykh_static_structure_factor}, which suggests that at $T=0.01 J$, the Heisenberg chain hosts 5-partite entanglement, with diverging entanglement depth at lower temperatures. By including $\epsilon(T)$ to compute $f_Q = 4S(\pi)-\epsilon$, we see that the chain actually hosts at-least $6$-partite entanglement at this temperature. We note that the Tomonaga-Luttinger-Liquid is a continuum field theory, which implies that the scaling form for the equal-time correlation function in Ref. \onlinecite{starykh1} is valid at distances $x> \Lambda \gg a$ where $\Lambda$ is a coarse-grained cutoff much larger than the lattice spacing $a$. Contributions from short distances $x\ll \Lambda$ to the true $S(\pi)$ will diverge as $T\rightarrow 0$ \cite{starykh2}. However, the scaling-form for the equal-time correlation function in Ref. \onlinecite{starykh1} produces finite contributions to $S(\pi)$ at short distances as $T\rightarrow 0$. The positivity of $S(\pi)$ then implies that the static structure factor from \eqref{starykh_static_structure_factor} is an underestimate, as it does not include the effect of diverging short distance contributions close to criticality, while capturing correlations at distances greater than the coarse-grained cutoff scale accurately. Together, these effects suggest that \eqref{starykh_static_structure_factor} produces at most a lower bound to the true entanglement depth at low $T$.

\subsection{\label{mps} Numerical Results}
\begin{figure}\label{numerics}
\centering
\includegraphics[width=8.6cm]{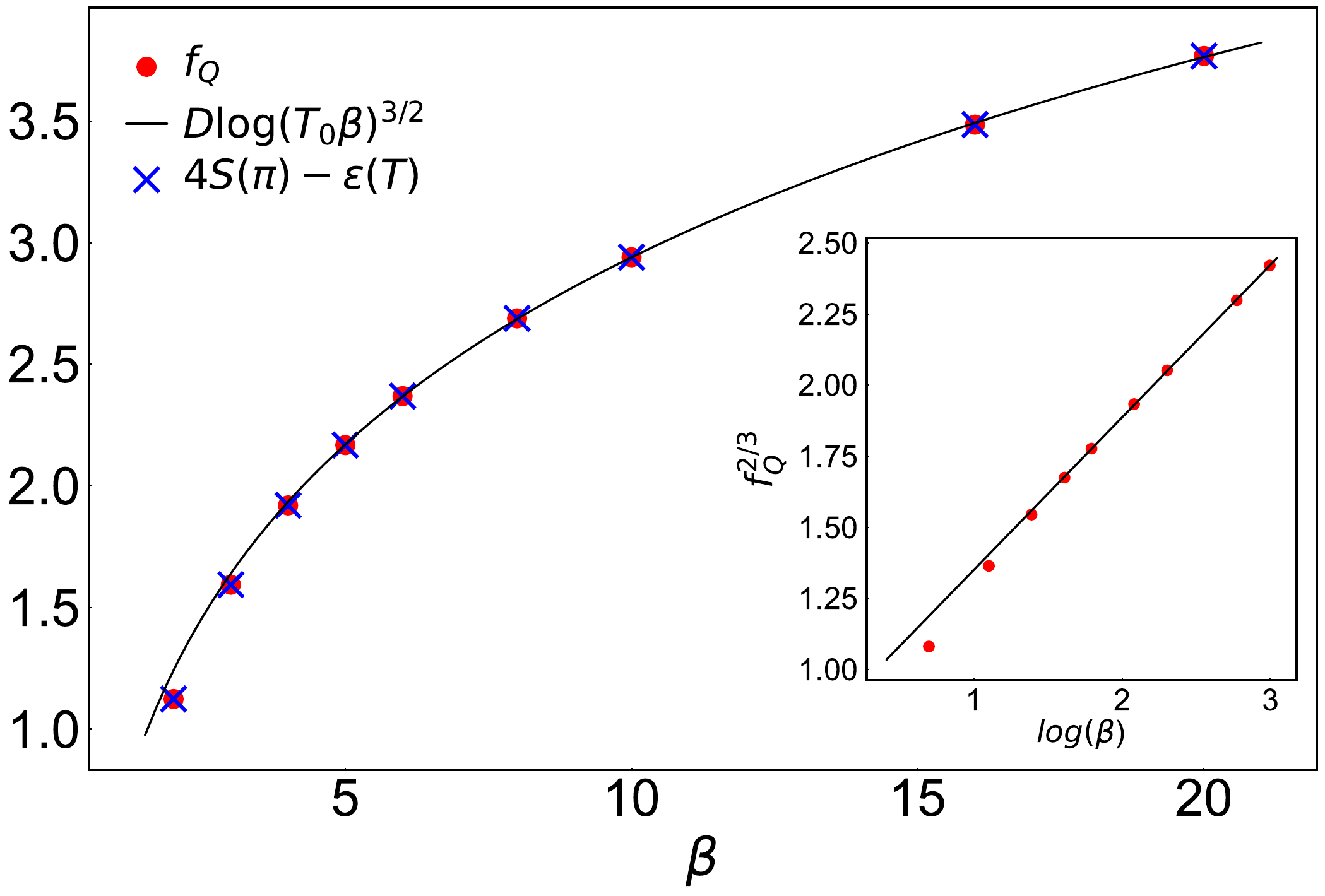}
\caption{$f_Q$ computed from MPS simulations of $S(\pi, \omega)$ using Eq. \eqref{qfi_integral}. The QFI density is fit to the scaling form $D\log(T_0\beta)$ from Eq. \eqref{static_heisenberg_scaling}, for $\beta \geq 4$. Crosses indicate $4S(q) - \epsilon$ computed from $S(\pi, \omega)$ using Eq. \eqref{dsf_sum_rule} and Eq. \eqref{QFI_correction}. Inset: Asymptotic scaling collapse of $f_Q^{2/3}$ against $\log(\beta)$, fit to $\beta \geq 4$.}
\end{figure}
We evaluate the QFI in the Heisenberg model and verify its asymptotic scaling from MPS simulations of the dynamic structure factor, on a chain of length $L=256$ with bond dimension $\chi=1024$. More details of the simulations are discussed in the appendix. \\

The inset of Fig. \ref{numerics} shows that the finite temperature QFI scales as we expect as $\log(\beta)^{3/2}$ in the low temperature limit. $f_Q^{2/3}$ is fit linearly against $\log(\beta)$ for $\beta \geq 4$, showing an asymptotic scaling collapse consistent with Eq. \eqref{QFI_scaling_T}. Fig. \ref{numerics} also shows excellent agreement with the power-logarithmic model in Eq. \eqref{static_heisenberg_scaling} even at intermediate temperatures. By extrapolating the QFI from the fit in Fig. \ref{numerics}, we find that $f_Q(T=0.01) \approx 5.9$, which suggests 6-partite entanglement at $T=0.01$, consistent with the prediction from the low-energy CFT methods in section \ref{CFT}. Moreover, we see that (bipartite) entanglement persists up to temperatures as high as $\beta = 2$ ($T=0.5 J$). In Fig. \ref{numerics}, we also show that $f_Q$ computed directly from equation \eqref{qfi_integral} exactly matches the QFI computed as $4S(\pi)-\epsilon$, where $S(\pi)$ is computed from MPS data for $S(\pi, \omega)$ using the sum-rule \eqref{dsf_sum_rule}, and $\epsilon$ is calculated using the integral formula \eqref{correction}, consistent with the decomposition of the QFI in Eq. \eqref{QFI_correction}.

\section{\label{experiment} Experimental Considerations}
\begin{figure}
\includegraphics[width=8.6cm]{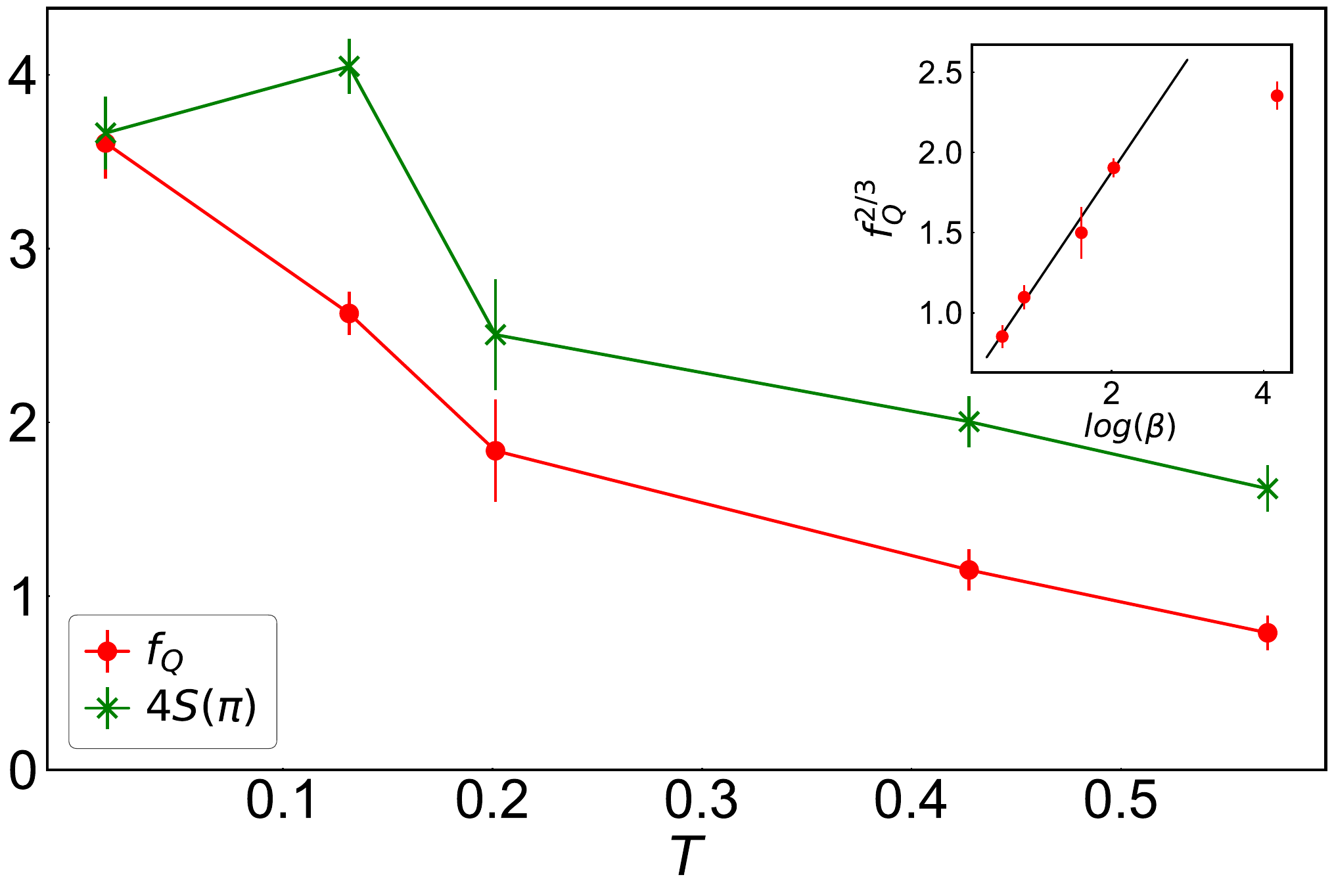}
\caption{Calculated normalized Quantum Fisher Information $f_Q$ and static structure factor $4S(\pi)$ for KCuF$_3$. Temperature is in units of $J$. Error bars indicate one standard deviation uncertainty. Inset: $f_Q^{2/3}$ shown to scale linearly with $\log(\beta)$ within uncertainty. The lowest temperature point is excluded from scaling due to non-negligible interchain coupling causing deviations from the one-dimensional model.}
	\label{fig:KCuF3}
\end{figure}
To experimentally demonstrate the relationship between the Quantum Fisher Information and integrated scattering, we calculate $f_Q$ and $4S(\pi)$ for the 1D Heisenberg spin chain KCuF$_3$, for which $J \approx 33.5$ meV, using the inelastic neutron scattering data from Ref. \cite{scheie_kcuf3}, shown in Fig. \ref{fig:KCuF3}. The measured normalized QFI in the temperature range of the experimental data is consistent with the predicted entanglement depths from the CFT theory and MPS simulations in sections \ref{CFT} and \ref{mps}. The inset of Fig. \ref{fig:KCuF3} shows the expected scaling collapse of $f_Q^{2/3}$ against $\log(\beta)$, where we exclude the lowest temperature point ($6\text{K} = 0.015 J$) from the fit because non-zero interchain coupling 
($J'/J = 0.047$ \cite{Lake_2005}) causes magnetic order below $39$K in KCuF$_3$, causing deviations from the idealized one-dimensional Heisenberg chain. For every temperature except the lowest ($6$K), $4S(\pi)$ is greater than $f_Q$ by approximately 1, confirming that total scattering is a reliable approximation to the QFI. At $6$K, $f_Q$ and $S(\pi)$ are identical to within uncertainty. This is partially because $4S(\pi)$ and $f_Q$ converge at the lowest temperatures, but also because KCuF$_3$ does not have a true logarithmic divergence at $6$K. The interchain coupling produces a finite energy maximum in KCuF$_3$ \cite{Lake_2000}, and thus the $\tanh$ factor in the integral form of the QFI \eqref{qfi_integral} suppresses a negligible amount of scattering at the lowest temperatures. Measuring other antiferromagnetic Heisenberg materials with smaller interchain coupling, such as CuPzN ($J'/J < 10^{-4}$) \cite{Hammar_1999} or Sr$_3$CuO$_3$ ($T_N \approx 0.002 J = 4.4K$) \cite{Motoyama_1996}, would allow $T_Q$ to be reached without magnetic order. Examining these materials in detail should yield multipartite entanglement far greater than KCuF$_3$. A noteworthy example to be considered for future experiments is Sr$_3$CuO$_3$, which has $J\approx190$ meV \cite{Motoyama_1996}. At $T=6 \ \text{K} \approx 0.0027 \ J$, the logarithmic fit in Fig. \ref{numerics} predicts $f_Q\approx 7.7$ -- an entanglement depth of at least $8$ for this material. \\

Another important consideration is that magnetic Bragg peaks in the scattering spectrum must be excluded from $S(q)$. The QFI explicitly excludes elastic scattering at $\omega = 0$ \cite{hauke}, and so must $S(q)$ if it is to be used as an entanglement witness. This is not a consideration if the material is being studied above its magnetic ordering temperature, however, below the ordering temperature, Bragg peaks develop which will increase the elastic contribution to $S(q)$ dramatically -- but this does not indicate increased entanglement. Thus $S(q)$ must be measured either (i) above the magnetic ordering temperature, (ii) at a reciprocal space vector transverse to chains where no Bragg intensity appears (as was done for KCuF$_3$ in Fig. \ref{fig:KCuF3}), or else (iii) the Bragg intensity must be manually removed. It should be noted that nonzero interchain coupling is unavoidable in real materials, and dimensional crossovers, where the system acts three-dimensional rather than one-dimensional, are manifest even above the ordering temperatures \cite{Dupont_2018,Lake_NM_2005}. Nevertheless, while such effects will cause deviation from theory and simulations, the QFI still gives a meaningful lower bound on multipartite entanglement. In fact, as the KCuF$_3$ example shows, without the true low-energy divergence, $4S(\pi)$ and $f_Q$ will converge even faster.

\section{\label{conclusion} Conclusion}
\subsection{Summary}
We studied the finite temperature multipartite entanglement properties of the 1-D spin-$\frac{1}{2}$ antiferromagnetic Heisenberg model, showing that multipartite entanglement scales as $\sim (\log(1/T))^{3/2}$. We also make an analytical argument that static structure factors of  certain operators witness entanglement below a characteristic temperature scale, by demonstrating a general decomposition for the Quantum Fisher Information (QFI) in terms of the static structure factor and a correction term that vanishes at $T\rightarrow 0$. Using these results, we determine the multipartite entanglement depth at low temperatures in the Heisenberg chain, and verify the logarithmic scaling law using results from conformal field-theory \cite{starykh1, starykh2} and MPS simulations, showing that entanglement persists at temperatures as high as $T = 0.5 J$. We also demonstrate agreement of our results with neutron scattering data for the Heisenberg-like material KCuF$_3$. Our work shows that the Heisenberg chain hosts a non-trivial logarithmic critical scaling of low-temperature multipartite entanglement, a result that complements the known logarithmic scaling of entanglement entropy at the Heisenberg critical point. Moreover, our results suggest that Heisenberg-like materials host high levels of entanglement even at intermediate temperatures, potentially useful for quantum technologies that require robust entanglement.

\subsection{Outlook}
A natural extension of this work would be to study the quantum critical scaling of multipartite entanglement at finite temperature in two-dimensional systems with dominant logarithmic corrections to spectral functions. One such example is the square-lattice spin-$\frac{1}{2}$ antiferromagnetic Heisenberg chain, for which the static structure factor and susceptibility have been studied numerically with Quantum Monte-Carlo simulations, experimentally in neutron scattering studies, as well as with chiral perturbation theory \cite{2D_Heisenberg, 2D_exp}. Another interesting direction could be to study the transition from logarithmic divergence of finite temperature multipartite entanglement to a power-law divergence as next-nearest neighbor interactions are tuned in the 1D Heisenberg chain, as sufficiently strong next-nearest neighbor interactions are expected to remove marginally irrelevant operators from the Bosonized Hamiltonian \cite{starykh1, starykh2}. Moreover, our work introduces many potential directions for experimental work to study multipartite entanglement at finite temperature, by measuring static structure factors in both 1D and 2D quantum materials, such as the 1D Heisenberg antiferromagnet Sr$_3$CuO$_3$ \cite{Motoyama_1996}, or the 2D square-lattice Heisenberg antiferromagnet Sr$_2$CuO$_2$Cl$_2$ \cite{2D_exp}.

\section{\label{acknowledgements} Acknowledgements}
This work was supported by National Science Foundation grant DMR-1918065 (V.M. and J.E.M.), and the U.S. Department of Energy, Office of Science, Office of Basic Energy Sciences Award No. DE-AC02-05-CH11231 through the Theory Institute for Materials and Energy Spectroscopies (TIMES) (N.S.). A.O.S. and D.A.T. were supported by the U.S. Department of Energy, Office of Science, National Quantum Information Sciences Research Centers, Quantum Science Center.
\appendix*

\section{Details of MPS Simulations}\label{details_numerics}
\begin{figure}
\includegraphics[width=8.6cm]{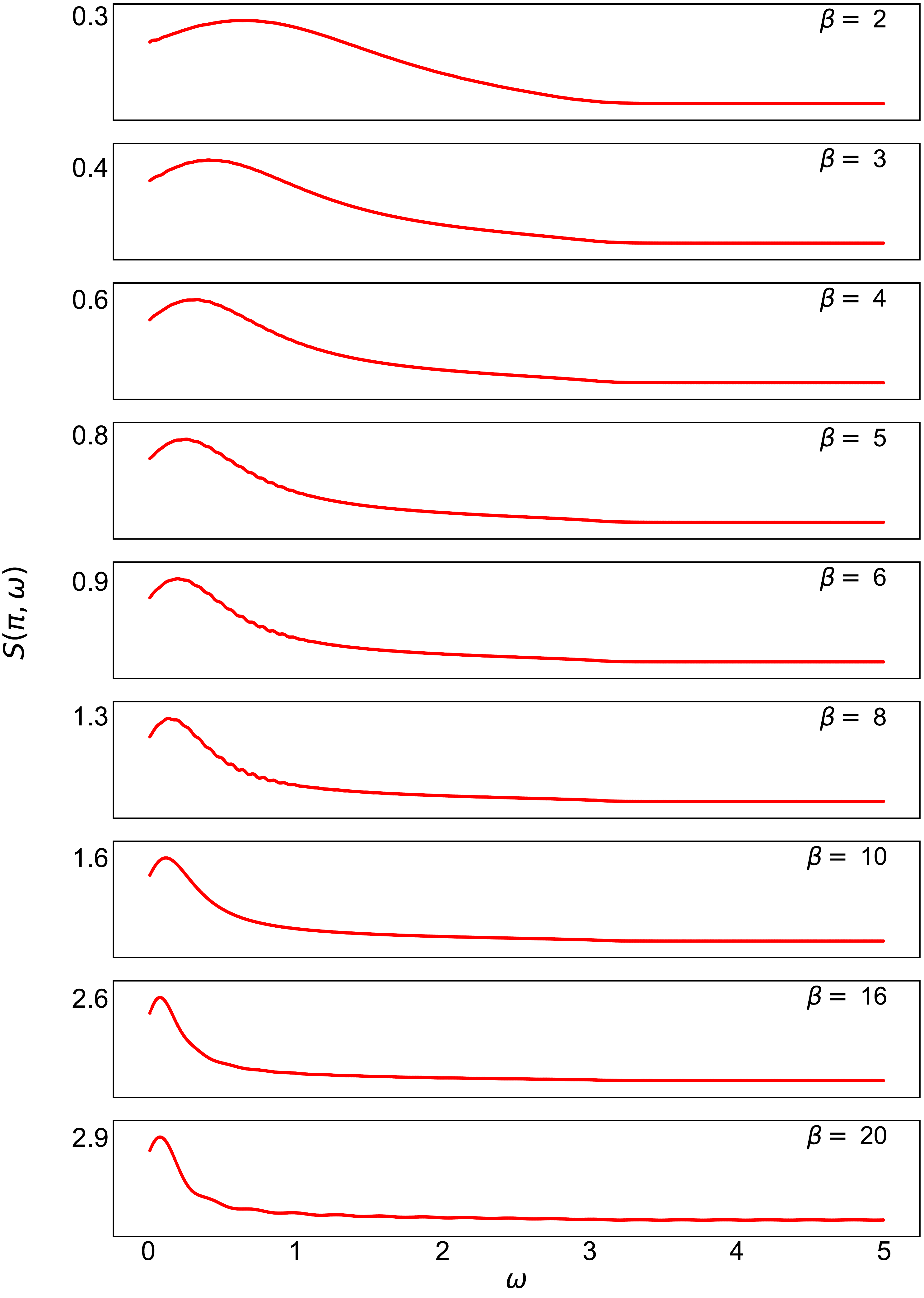}
\caption{\label{DSF_allT} Finite temperature staggered spin dynamic structure factor computed from MPS and TEBD simulations for $\beta = 4,5,6,8,10,16,20$.}
\end{figure}

In section \ref{numerics}, we calculate the QFI from the dynamic structure factor according to Eq. \eqref{qfi_dsf}, which is computed from MPS simulations of two-point space-time correlation functions of the $S^z$ operator, 
\begin{equation*}
    G(x,t) = <S^z_x(t), S^z_c(0)>
\end{equation*}
where $c = L/2$ is the center site of the chain. We use data for the two-point correlation functions from a previous work \cite{maxime}, computed from matrix product states of length $L=256$, with bond dimension $\chi=1024$. For the time dependence, we use the Time Evolution Block Decimation (TEBD) \cite{schollwock_mps} algorithm with $\Delta t = 0.1$ and a maximum time of $t_{max} = 100$. This in turn sets a minimum and maximum range of confidence in the $\omega$ dependence of $S(q,\omega)$ between $\omega_{min} \sim \frac{1}{t_{max}}$ and $\omega_{max} \sim \frac{1}{\Delta t}$. For the finite temperature dependence, we use the ancilla purification method along with imaginary time evolution \cite{schollwock_mps,finite_T_DMRG} with an ancilla system of equal length, and $\Delta \beta = 0.1$ for the imaginary time block decimation. As shown in our previous work, the correlation functions have been checked for convergence in $L$, $\chi$, and $t_{max}$ \cite{maxime}. Using space-time translational invariance, the dynamic structure factor is defined as 
\begin{equation*}
        S(q,\omega) =  \frac{1}{2\pi N}\int_{-\infty}^\infty \mathrm{d}t \sum\limits_{x}e^{-i(q(x-c)-\omega t)}G(x,t)
\end{equation*}
and is a real and non-negative quantity. By exploiting symmetry relations of the equal time correlation function \cite{recursion}, we may write the DSF in terms of only the positive time correlations as 
\begin{align*}
        S(q,\omega) &= \frac{1}{\sqrt{N}}\int_{0}^\infty \mathrm{d}t \sum\limits_{x}\cos(q(x-c)) \\
        &\times \left(\cos(\omega t)\mathrm{Re}(G(x,t))-\sin(\omega t)\mathrm{Im}(G(x,t))\right)
\end{align*}
which additionally ensures that $S(q,\omega)$ is strictly real \cite{scheie}. We then normalize $S(q,w)$ so that it satisfies the total inelastic moment sum rule for one spin component in the isotropic Heisenberg model \cite{recursion}:
\begin{equation*}
    \int_{-\infty}^{\infty}\int_{0}^{2\pi}\mathrm{dq}\ \mathrm{d\omega}\ S^z(q,\omega) = \frac{S(S+1)}{3} = \frac{1}{4}
\end{equation*}
where $S=\frac{1}{2}$ is the spin of the model, and the $z$ superscript on $S^z(q,\omega)$ is to explicitly indicate that this sum rule holds for one spin component of the full DSF. This gives $S^z(q,\omega)$ in units of $2\pi$, so we renormalize $S^z(q,\omega)$ accordingly to absolute units. To mitigate the effects of finite spectral resolution, we use Gaussian broadening when computing the DSF by making the substitution
\begin{equation*}
    G(x,t) \rightarrow G(x,t)e^{-\eta t^2}
\end{equation*}
We use the smallest value of $\eta$ that produces a strictly non-negative spectrum. This amounts to $\eta\approx 0.01$ for all $T$ in our simulations. In Fig. \ref{DSF_allT}, we show the resulting spectra for the staggered component of the DSF, $q=\pi$, from of our data analysis at inverse temperatures $\beta = 4,5,6,8,10,16,20$ that is used to calculate the QFI values in Fig. \ref{numerics}. The maximum value of $\beta$ in our range of data is $\beta = 20$. Since the dynamical critical exponent for the Heisenberg universality class is $z=1$, finite temperature introduces a cutoff scale in the correlation length which is at most $\lambda \sim \beta^{1/z} = 20 \ll 256$. Similarly, using finite entanglement scaling arguments \cite{finite_entanglement}, it can be shown that the length scale introduced by the finite bond dimension $\chi = 1024$ is much larger than $\beta^{1/z} = 20$. Thus, finite size and finite bond dimension effects are expected to be negligible for the majority of the spectrum, as temperature is the most relevant perturbation away from criticality. However, as $\omega \rightarrow 0$, $S(\pi,\omega)$ diverges as a power law at low temperature \cite{Muller,Tennant}, and effects from finite $t_{max}$, size, and bond dimension can cause the tensor network description to underestimate the critical divergence at $\omega=0$ as $T\rightarrow 0$. Thus, the QFI computed from our MPS simulations is also a lower-bound in the worst case.\\

\newpage
\bibliography{apssamp}

\end{document}